\documentclass[aps,preprint,epsfig,rotate]{revtex4}
\usepackage{graphicx}
\usepackage{bm}
\usepackage{epsfig}

\input epsf
\begin{document}
\title{Atomic fragments from the nuclear reaction of the ${}^{6}$Li atom
       with slow neutrons}

 \author{Alexei M. Frolov}
 \email[E--mail address: ]{afrolov@uwo.ca}

 \author{David M. Wardlaw}
 \email[E--mail address: ]{dwardlaw@uwo.ca}

\affiliation{Department of Chemistry\\
 University of Western Ontario, London, Ontario N6H 5B7, Canada}

\date{\today}

\begin{abstract}

Approximate probabilities of formation of various atoms and ions in
different bound states are determined for the exothermic nuclear
$(n,{}^{6}$Li$;t,{}^{4}$He)-reaction of atomic lithium-6 with slow
neutrons. In our calculations of the final state probabilities we have
assumed that the incident lithium atom is in its ground (doublet) atomic
${}^2S(L = 0)-$state. It is straightforward to generalize our analysis to
other bound states of the three-electron Li atom.

\end{abstract}
\maketitle
\newpage

\section{Introduction}

The nuclear reaction of ${}^{6}$Li nuclei with slow neutrons is written in
the form \cite{Kik}
\begin{eqnarray}
 {}^6{\rm Li} + n = {}^4{\rm He} + t + 4.785 \; \; MeV \label{e1}
\end{eqnarray}
where the notations ${}^4$He and $t$ stand for the helium nucleus (often
also called the $\alpha-$particle, or $\alpha$, for short) and tritium (or
${}^3$H) nucleus. For thermal neutrons with $E_n \approx 0$ the
cross-section $\sigma$ of this nuclear reaction is very large $\sigma_{max}
\approx 960 \cdot 10^{-24}$ $cm^2$ or 960 $barn$ \cite{WW}, for short. The
velocities of the two nuclear fragments formed in the reaction,
Eq.(\ref{e1}), with slow neutrons are $v_t \approx$ 6.03986 $a.u.$ and
$v_{\alpha} \approx$ 4.52989 $a.u.$ for the tritium nucleus and
$\alpha-$particle, respectively. These velocities are given in atomic units,
where $\hbar = 1, m_e = 1, e = 1$. The unit of atomic velocity is $v_e =
\alpha c \approx \frac{c}{137} \approx 2.1882661 \cdot 10^{8}$ $cm \cdot
sec^{-1}$, where $c$ is the speed of light and $\alpha$ is the fine
structure constant. Formally, the atomic velocity $v_e$ is the velocity of
the $1s-$electron in the hydrogen atom with the infinitely heavy nucleus
${}^{\infty}$H. It is clear that in atomic units $v_e = 1$.

The nuclear reaction of the ${}^{6}$Li nucleus, Eq.(\ref{e1}), proceeds with
neutrons of all energies and the energy released increases almost linearly
with the energy of the incident neutron. For fast neutrons with $E_n \ge 1$
$MeV$ the energy released is substantially different from the value quoted
in Eq.(\ref{e1}) and the velocities of the atomic fragments increase
correspondingly. Note that the cross-section of this reaction has a large
resonance (maximum) $\sigma \approx 4.5$ $barn$ at $E_n \approx 240 - 270$
$keV$, but it is relatively large for neutrons of all energies $E_n \le 0.8$
$MeV$. This makes the reaction, Eq.(\ref{e1}), extremely important in the
thermonuclear ignition and following propagation of the thermonuclear
burning wave in highly compressed ($\rho \ge 100$ $g \cdot cm^{-3}$)
${}^6$LiD deuteride which is routinely used as a thermonuclear fuel in
modern thermonuclear explosive devices (see, e.g., \cite{Gon},
\cite{Fro98}). The analogous ($n, {}^{3}$He)-reaction \cite{FrWa09} also
plays a very important role in such processes.

In general, the nuclear $(n,t)$-reactions of the ${}^{6}$Li and ${}^{3}$He
nuclei with neutrons allow one to: (1) reduce drastically the overall
bremsstrahlung loss from the combustion zone; and (2) increase the
tritium/deuterium ratio which is crucially important to start many new
thermonuclear $(d,t)-$reactions. Briefly, by using the ${}^{6}$LiD deuteride
in modern thermonuclear explosive devices we can reduce the required
compressions to relatively small values. In many cases such compressions are
dozens times smaller (usually, in 25 - 40 times smaller \cite{Fro98}) than
compressions required for any other (solid) thermonuclear fuel, e.g.,
${}^{7}$LiD deuteride. On the other hand, by using compressions which are
provided by a standard `primary' nuclear charge one can create extremely
compact thermonuclear explosive devices based on the ${}^{6}$LiD deuteride.
The idea to use the pure ${}^{6}$LiD deuteride in thermonuclear explosive
devices was originally proposed by V.L. Ginzburg in 1948-1949 (see
discussion and references in \cite{Gon}).

Our goal in this study is to make accurate numerical predictions of the
final state probabilities for the reaction, Eq.(\ref{e1}), with slow
neutrons. We want to predict (accurately) the probabilities of formation of
various atoms and ions in different (ground and excited) final states.
Briefly, we want to determine the probabilities to detect the He and
${}^{3}$H atoms and He$^{+}, {}^{3}$H$^{-}$ ions in their bound states. Note
that all newly created atomic fragments from reaction, Eq.(\ref{e1}), move
rapidly even in the case of slow neutrons and that accurate theoretical
prediction of the final state probabilities for rapidly moving atomic
fragments is not trivial.

On the other hand, the results of our evaluations and methods
created for such evaluations are of great interest in many applications
related with the nuclear reaction, Eq.(\ref{e1}), and another similar
reaction of ${}^{10}$B nuclei with slow neutrons, namely
\begin{equation}
 {}^{10}{\rm B} + n = {}^7{\rm Li} + {}^4{\rm He} + 2.791 \; \; MeV
 \label{e2}
\end{equation}
The reaction, Eq.(\ref{e2}), is extensively used in the boron neutron
capture therapy (BNCT, for short), or boron neutron-capture synovectomy
\cite{BNCT1} - \cite{Haw3}, to treat different forms of cancer, including
brain cancer. The fast $\alpha-$particle produced in the reaction,
Eq.(\ref{e2}), kills (or at least `badly damages') one cancer cell before
it finally stops. The modern applications of this reaction to cancer
treatment are based on the use of molecules which contain a large number of
${}^{10}$B-atoms, e.g., the Na$_3$[B$_{20}$ H$_{17}$NH$_{2}$
CH$_{2}$CH$_{2}$ NH$_{2}$] molecule, other similar molecules, and molecular
clusters \cite{Haw1}, \cite{Haw2} (see also \cite{Haw3} and references
therein). In this case the overall energy release from the reaction,
Eq.(\ref{e2}), in one cancer cell can be extremely large. Correspondingly,
the local temperature in the whole cell suddenly increases to very large
values and this kills the incident cancer cell with almost 100 \%
probability.

Note that the tritium nucleus does not form in the nuclear reaction
Eq.(\ref{e2}) which means that it is safe to initiate this reaction inside
of a human body. By studying the nuclear reaction, Eq.(\ref{e1}), in
few-electron atoms and ions we want to develop a number of reliable
theoretical methods and numerical procedures which can be later used in
applications to the analogous reaction, Eq.(\ref{e2}).

\section{Approximate variational wave function of the lithium atom}

In our earlier paper \cite{FrWa09} we have calculated the final state 
probabilities to form one-electron atomic species in those cases when 
exothermic nuclear $(n;t)-$ and $(n;\alpha)-$reactions occur in one 
electron atoms and ions. In this study we deal with the actual 
three-electron wave function of the Li atom and two-electron wave function 
of the He atom. For our purposes in this  study it is important to 
construct accurate variational wave function(s) of the ground (doublet) 
${}^2S(L = 0)-$state of the Li atom which is written in the following 
general form (see, e.g., \cite{Fro2011}, \cite{Lars})
\begin{eqnarray}
 \Psi({\rm Li})_{L=0} = \psi_{L=0}(A; \bigl\{ r_{ij} \bigr\}) (\alpha \beta
 \alpha - \beta \alpha \alpha) + \phi_{L=0}(B; \bigl\{ r_{ij} \bigr\}) (2
 \alpha \alpha \beta  - \beta \alpha \alpha - \alpha \beta \alpha)
 \label{psi}
\end{eqnarray}
where $\psi_{L=0}(A; \bigl\{ r_{ij} \bigr\})$ and $\phi_{L=0}(B; \bigl\{
r_{ij} \bigr\})$ are the two independent radial parts (= spatial parts) of
the total wave function. Everywhere below in this study, we shall assume
that all mentioned wave functions have unit norm. The notations $\alpha$ and
$\beta$ in Eq.(\ref{psi}) are the one-electron spin-up and spin-down
functions, respectively (see, e.g., \cite{Dir}). The notations $A$ and $B$
in Eq.(\ref{psi}) mean that the two sets of non-linear parameters associated
with radial functions $\psi$ and $\phi$ can be optimized independently. Note
that each of the radial basis functions in Eq.(\ref{psi}) explicitly depends
upon all six interparticle (or relative) coordinates $r_{12}, r_{13},
r_{23}, r_{14}, r_{24}, r_{34}$, where the indexes 1, 2, 3 stand for the 
three electrons, while index 4 mean the nucleus.
 
In our earlier work \cite{Fro2010} we have introduced an advanced set of
radial basis functions for bound state computations of three-electron atomic
systems. Such a set is called the semi-exponential basis set and it is
written in the form
\begin{eqnarray}
 \psi_{L=0}(A; \bigl\{ r_{ij} \bigr\}) = \sum^N_{k=1} C_k r^{n_1(k)}_{23}
 r^{n_2(k)}_{13} r^{n_3(k)}_{12} r^{m_1(k)}_{14} r^{m_2(k)}_{24}
 r^{m_3(k)}_{34} exp(-\alpha_{k} r_{14} -\beta_{k} r_{24} -\gamma_{k}
 r_{34}) \label{semexp}
\end{eqnarray}
where $\alpha_k, \beta_k, \gamma_k$ ($k = 1, 2, \ldots, N$) are the varied
non-linear parameters. The use of a large number of non-linear parameters
in Eq.(\ref{semexp}) allows one to construct compact and very accurate
variational wave functions for different three-electron atoms and ions. It
was shown in \cite{Fro2011} that the semi-exponential basis,
Eq.(\ref{semexp}), has a large number of other advantages in accurate
numerical computations.

In the sudden approximation \cite{Mig1}, \cite{Mig2} the numerical
determination of the final state probabilities is reduced to the analytical
computation of the Fourier transform of the overlap integral between the
incident $\psi_{in}$ and final $\psi_{fi}$ wave functions. In the case of
the nuclear reaction of ${}^{6}$Li nuclei with slow neutrons under
consideration here, the incident wave function represents the ground
${}^2S-$state of the lithium atom. The final wave function represents the
product atom and/or ion. For the helium atom product we need to determine
the following integral (or probability amplitude $M_{if}$)
\begin{eqnarray}
 M_{if} = \int \int \int \Psi_{\rm Li}(r_{14}, r_{24}, r_{34}, r_{12},
 r_{13}, r_{23}) \exp(\imath {\bf V}_{\alpha} \cdot {\bf r}_{14} +\imath
 {\bf V}_{\alpha} \cdot {\bf r}_{24}) \times \label{intg1} \\
 \Psi_{\rm He}(r_{14}, r_{24}, r_{12}) d^{3}{\bf r}_{14} d^{3}{\bf r}_{24}
 d^{3}{\bf r}_{34} \nonumber
\end{eqnarray}
where $\Psi_{\rm Li}$ is the wave function of the Li atom, while $\Psi_{\rm
He}$ is the wave function of the final (bound) state of the He atom. The
velocity ${\bf V}_{\alpha}$ is the final velocity of the $\alpha-$particle
after reaction, Eq.(\ref{e1}). In reality, we need to determine the 
correspponding probabilities $P_{if} =  M^{*}_{if} \cdot  M_{if}$. This 
allows one to reduce the problem (see below) to numerical computation of 
the one- and two-electron density matrix.   

For the tritium atom product in one of its bound states the probability 
amplitude $M_{if}$ is written in a slightly different form
\begin{eqnarray}
 M_{if} = \int \int \int \Psi_{\rm Li}(r_{14}, r_{24}, r_{34}, r_{12},
 r_{13}, r_{23}) \exp(\imath {\bf V}_{t} \cdot {\bf r}_{14})
 \Psi_{\rm H}(r_{14}) d^{3}{\bf r}_{14} d^{3}{\bf r}_{24}
 d^{3}{\bf r}_{34} \label{intg2}
\end{eqnarray}
where $\Psi_{\rm H}(r_{14})$ is the unit-norm wave function of the tritium
(or hydrogen) atom and ${\bf V}_{t}$ is its velocity after the reaction,
Eq.(\ref{e1}). In our earlier study \cite{FroWar2011} we have calculated
integrals similar to the integrals in Eq.(\ref{intg1}) and Eq.(\ref{intg2})
for the two-electron atomic systems which are involved in the analogous
nuclear reaction of the ${}^{3}$He nuclei with slow neutrons. The integral,
Eq.(\ref{intg1}), can be considered as a partial Fourier transformation of
the overlap of the incident and final wave functions. This integral
represents the Galilean transformation between the incident system (which
was at rest) and the final system which is rapidly moving with constant
speed.

In general, numerical calculations of integrals Eqs.(\ref{intg1}) and
(\ref{intg2}) are difficult to perform due to the presence of the
electron-electron coordinates $r_{ij}$ in the wave function. On the other
hand, for approximate evaluation of the final state probabilities for the
nuclear reaction, Eq.(\ref{e1}), we do not need to use the highly accurate
wave functions of the Li and He atoms. In particular, we can use the wave
functions of few-electron atomic systems which do not contain any of the
electron-electron coordinates. This drastically simplifies
analytical/numerical calculation of the integrals, Eq.(\ref{intg1}) -
Eq.(\ref{intg2}). Formally, such approximate wave functions depend on the
electron-nucleus coordinates only. Therefore, these wave functions
correspond to a system of $A-$electrons which do not interact with each
other and can be considered as free-electron wave functions. For the
He atom we have $A = 2$, while for the Li atom we have $A = 3$. The
explicit construction of the approximate free-electron wave functions for
few-electron atomic systems is discussed below.

\subsection{Special form of the trial wave function of the lithium atom}

To avoid problems related to the analytical computation of the Fourier
transformations given by Eq.(\ref{intg1}) one can apply approximate
variational expansions of the three-electron wave function. To simplify all
following calculations we represent the wave function of the Li atom in
the form
\begin{eqnarray}
 \psi_{L=0}(r_{14}, r_{24}, r_{34}, 0, 0, 0) = \sum^{N_s}_{k=1}
 C_k r^{m_1(k)}_{14} r^{m_2(k)}_{24} r^{m_3(k)}_{34}
 exp(-\alpha_{k} r_{14} -\beta_{k} r_{24} -\gamma_{k} r_{34})
 \label{exp1} \\
 = \sum^{N_s}_{k=1} C_k r^{m_1(k)}_{1} r^{m_2(k)}_{2} r^{m_3(k)}_{3}
 exp(-\alpha_{k} r_{1} -\beta_{k} r_{2} -\gamma_{k} r_{3}) \nonumber
\end{eqnarray}
where $C_k$ are the linear (or variational) coefficients, while $m_1(k),
m_2(k)$ and $m_3(k)$ are the three integer (non-negative) parameters, which
are, in fact, the powers of the three electron-nucleus coordinates $r_{i4}
= r_i$ ($i$ = 1, 2, 3). Below, we shall assume that the trial wave function
Eq.(\ref{exp1}) has a unit norm. Furthermore, in all calculations performed
for this study only one spin function $\chi_1 = \alpha \beta \alpha - \beta
\alpha \alpha$ is used. It is clear that the wave function Eq.(\ref{exp1})
contains only electron-nuclear coordinates and does not include any of the
electron-electron coordinates. The real (and non-negative) parameters
$\alpha_{k}, \beta_{k}, \gamma_{k}$ are the $3 N_s$ varied parameters of the
variational expansion, Eq.(\ref{exp1}). The wave function, Eq.(\ref{exp1}),
must be properly symmetrized upon all three electron coordinates. This
problem is discussed in the next Subsection.

The principal question for the wave function, Eq.(\ref{exp1}), is related to
its overall accuracy. If (and only if) such accuracy is relatively high,
then such a wave function, Eq.(\ref{exp1}), can be used in actual
computations of the probability amplitudes, Eqs.(\ref{intg1}) and
(\ref{intg2}). In actual applications the approximate wave function,
Eqs.(\ref{intg1}) and (\ref{intg2}), can be constructed from the highly
accurate wave functions already known from earlier works (see, e.g.,
\cite{Fro2011} and references therein). Briefly, we can take our wave
function of the ground ${}^2S-$state of the Li atom from \cite{Fro2011} and
remove all those terms which contain electron-electron $r_{ij}$ coordinates.
Then the non-linear parameters in the trial wave function, Eq.(\ref{exp1}),
must be re-optimized. The resulting wave function can be considered as an
optimal independent-electron wave function of the ground state of the
${}^{\infty}$Li atom. Using this approach we have determined the 23-term
variational wave function shown in Table I. The total energy $E$ of the
ground ${}^2S-$state of the ${}^{\infty}$Li atom obtained with this 
independent-electron wave function is -7.44859276608 $a.u.$ Note that this 
value of $E$ is close to the exact total energy of the ground state of the 
${}^{\infty}$Li atom. This indicates a very good overall quality for our 
approximate wave function which does not include any of the 
electron-electron coordinates $r_{12}, r_{13}, r_{23}$. Note also that in
atomic physics based on the `Hatree-Fock' and even `hydrogenic' 
approximations the ground state in the Li atom is designated as the 
$2^2S-$state, while in the classification scheme developed in highly  
accurate computations the same state is often designated as the 
$1^2S-$state. This classification scheme is very convinient to work with 
trully correlated few-electron wave functions. It is clear that no 
hydrogenic quantum numbers are good in such cases, and we have to use the 
more appropriate (and convinient) classification scheme. To avoid 
confllicts between these two classification schemes in this study we 
follow the system of notation used by Larsson \cite{Lars} which 
designated this state in the Li atom as the `ground ${}^2S$-state'. 

\subsection{Antisymmetrization of the trial wave function of the lithium
atom}

The actual many-electron wave function in an atomic system must be
completely antisymmetric with respect to all electron variables, i.e. upon
all electron spatial and spin variables. This statement is true for all
exact and approximate few-electron wave functions, including the optimized
free-electron wave functions, Eq.(\ref{exp1}). Antisymmetrization of the
two-electron wave function is trivial and is not discussed here. For a
three-electron atomic wave function this requirement is written in the form
${\hat{\cal A}}_{123} \Psi(1,2,3) = - \Psi(1,2,3)$, where $\Psi$ is given
by Eq.(\ref{psi}) and $\hat{{\cal A}}_e$ is the three-particle (= electron)
antisymmetrizer ${\hat{\cal A}}_e = \hat{e} - \hat{P}_{12} - \hat{P}_{13} -
\hat{P}_{23} + \hat{P}_{123} + \hat{P}_{132}$. Here $\hat{e}$ is the
identity permutation, while $\hat{P}_{ij}$ is the permutation of the $i$-th
and $j$-th particles. Analogously, the operator $\hat{P}_{ijk}$ is the
permutation of the $i$-th, $j$-th and $k$-th particles. In actual
computations antisymmetrization of the total wave function is reduced to the
proper antisymmetrization of corresponding matrix elements (for more detail,
see, e.g., \cite{Fro2010}). Each of these matrix elements is written in the
form $\langle \Psi \mid \hat{O} \mid \Psi \rangle$, where $\hat{O}$ is an
arbitrary spin-independent quantum operator which is symmetric upon
all interparticle permutations. The wave function $\Psi$, Eq.(\ref{psi}),
contains the two different radial parts $\psi$ and $\phi$. By performing the
integration over all spin coordinates one finds the four following spatial
projectors ${\cal P}_{\psi\psi}, {\cal P}_{\psi\phi} = {\cal P}_{\phi\psi}$
and ${\cal P}_{\phi\phi}$
\begin{eqnarray}
 {\cal P}_{\psi\psi} = \frac{1}{2 \sqrt{3}} \Bigl( 2 \hat{e} + 2
 \hat{P}_{12} - \hat{P}_{13} - \hat{P}_{23} - \hat{P}_{123} - \hat{P}_{132}
 \Bigr) \\
 {\cal P}_{\psi\phi} = \frac12 \Bigl( \hat{P}_{13} - \hat{P}_{23} +
 \hat{P}_{123} - \hat{P}_{132} \Bigr) \\
 {\cal P}_{\phi\psi} = \frac12 \Bigl( \hat{P}_{13} - \hat{P}_{23} +
 \hat{P}_{123} - \hat{P}_{132} \Bigr) \\
 {\cal P}_{\phi\phi} = \frac{1}{2 \sqrt{3}} \Bigl( 2 \hat{e} - 2
 \hat{P}_{12} + \hat{P}_{13} + \hat{P}_{23} - \hat{P}_{123} -
 \hat{P}_{132} \Bigr)
\end{eqnarray}
Here the indexes $\psi$ and $\phi$ correspond to the notations used in
Eq.(\ref{psi}) to designate the two spatial parts of the total wave
function. For an arbitrary symmetric spin-independent operator $\hat{O}$
each of these four projectors produces matrix elements $\langle \Psi \mid
\hat{O} \mid \Psi \rangle$ of the correct permutation symmetry (for doublet
states) between all three electrons.

\subsection{Bound state wave functions of the final atomic fragments}

The final atomic states arising in the exothermic nuclear $(n,{}^{6}$Li$;t,
{}^{4}$He)-reaction of atomic lithium-6 with slow neutrons contain either
one, or two, or zero bound electrons. In this study our main interest is
restricted to one- and two-electron atoms and ions. The explicit form
of one-electron atomic wave functions takes the form (see, e.g., \cite{LLQ})
$\Phi_{n \ell m}(r, \Theta, \phi) \alpha = R_{n \ell}(Q, r) Y_{\ell
m}(\Theta, \phi) \alpha$, where $\alpha$ is the spin-up wave function,
$Y_{\ell m}(\Theta, \phi) = Y_{\ell m}({\bf n})$ is a spherical harmonic and
$R_{n \ell}(Q, r)$ is the radial function. The radial function is written in
the form
\begin{eqnarray}
  R_{n \ell}(Q,r) = \frac{1}{r n} \sqrt{\frac{Q (n - \ell - 1)!}{(n +
  \ell)!}} \Bigl[ \frac{2 Q r}{n} \Bigr]^{\ell + 1} \sum^{n-\ell-1}_{k=0}
  \frac{(-1)^k}{k!}
  \left(
  \begin{array}{c}
   n + \ell \\
   2 \ell + k + 1
  \end{array}
  \right)
  \Bigl[ \frac{2 Q r}{n} \Bigr]^{k} \exp\Bigl(-\frac{Q r}{n}\Bigr)
  \label{hydr}
\end{eqnarray}
where $Q$ is the nuclear charge, while $n$ and $\ell$ are the quantum
numbers of this bound state. Note that the radial function, Eq.(\ref{hydr}),
has a unit norm for an arbitrary $Q$.

The wave function of the two-electron He-atom can be approximated in a
variety of different forms. Currently, it is possible to construct
approximate wave functions which provide 25 - 50 correct decimal digits for
the total non-relativistic energy of the He atom and He-like ions. In this 
study, however, we shall use an approximate He wave function which does not 
contain the electron-electron coordinate $r_{21}$. For the ground 
$1^1S-$state of the ${}^{\infty}$He atom the radial part of such a wave 
function is written in the form
\begin{eqnarray}
 \psi_{L=0}(r_1, r_2, r_{12}) = C \exp\Bigl[-(Q - \frac{5}{16}) r_1
 -(Q - \frac{5}{16}) r_2\Bigr] \label{heatom}
\end{eqnarray}
where $Q = 2$ is the nuclear charge for the He nucleus and $C = \frac{(Q 
- \frac{5}{16})}{2}$ is the normalization constant. The corresponding spin 
part of the total wave function takes the form $\eta = \alpha(1) \beta(2) - 
\beta(1) \alpha(2) = \alpha \beta - \beta \alpha$. The explicit form of the 
$\eta$ spin function is important in the performance of integration over all 
spin coordinates.

The total energy of the ${}^{\infty}$He atom obtained with this wave
function, Eq.(\ref{heatom}), is -2.85 a.u. which indicates that it provides
a relatively good approximation to the actual ground state wave function.
The main advantage of the approximate wave function, Eq.(\ref{heatom}), is
its explicit dependence upon the two electron-nuclear coordinates $r_1$ and
$r_2$ only. This drastically simplifies the subsequent computation of radial
integrals with the Bessel functions (see below). On the other hand, we need
to note that the best-to-date one-term radial wave function for the
${}^{\infty}$He atom (see, e.g., \cite{FrWa09}) corresponds to substantially
better numerical accuracy, since it provides the total energy $E$ = -2.899
534 375 443 69 $a.u.$ which is very close to the exact answer (see, e.g.,
\cite{Fro2007}). The corresponding non-linear parameters can be found in
\cite{FrWa09}. But, in contrast with the wave function, Eq.(\ref{heatom}),
the wave function from \cite{FrWa09} explicitly depends upon the
electron-electron coordinate $r_{21}$.

\section{Calculations}

By using the free-electron wave function of the Li atom obtained above we
can estimate the probabilities to form various atomic species during the
nuclear reaction, Eq.(\ref{e1}). The structure of the trial wave function,
Eq.(\ref{exp1}), enables one to perform accurate computations of all
integrals which include one, two and even three Galilean exponents for
electrons. In reality all such integrals are reduced to the products of
one-dimensional integrals. In other words, in our approach all electron
coordinates are separated and this simplifies drastically the following
analytical and numerical computations of all required integrals.

To determine the matrix element of the operator $\exp (\imath {\bf V} \cdot
{\bf r})$ (Galilean exponent) between the wave functions of the initial and
final bound states, we apply the Rayleigh expansion of a plane wave (with 
the use of spherical harmonic addition theorem \cite{LLQ}, \cite{Ros}). An 
alternative approach is based on direct calculations of the integrals 
containing the $\exp(-A r + \imath {\bf V} \cdot {\bf r})$ factor. The 
explicit formula takes the form (see, e.g., \cite{LLQ}, \cite{Ros}, 
\cite{Brin})
\begin{eqnarray}
 \exp (\imath {\bf V} \cdot {\bf r}) = 4 \pi \sum_{\ell=0} \imath^{\ell}
 j_{\ell}(V r) \sum^{\ell}_{m=-\ell} Y^{*}_{\ell m}({\bf n}_V)
 Y_{\ell m}({\bf n}_r) \label{ray}
\end{eqnarray}
where $Y_{\ell m}({\bf n})$ are the spherical harmonics, ${\bf V}$ is the
velocity of the final atomic fragment and ${\bf n}_y = \frac{{\bf y}}{y}$ is
the unit norm vector which corresponds to an arbitrary non-zero vector ${\bf
y}$. Also in this equation the spherical Bessel functions $j_{\ell}(V r)$
are defined by the relation (see, e.g., \cite{GR}, \cite{AS})
\begin{equation}
 j_{\ell}(V r) = \sqrt{\frac{\pi}{2 V r}} J_{\ell+\frac12}(V r) \label{eq2}
\end{equation}
where $J_{\ell+\frac12}(x)$ are the Bessel functions. These formulas are
used in analytical and/or numerical computation of all required matrix
elements.

Actual computations of matrix elements with the `factorized' trial wave
functions, Eq.(\ref{exp1}) are performed with the use of the following
formula
\begin{eqnarray}
 \int^{\infty}_{0} t^{\mu} J_{\nu}(b t) exp(-p t) dt =
 \frac{\Gamma(\mu + \nu + 1)}{\Gamma(\nu + 1)} \Bigl( \frac{b}{2}
 \Bigr)^{\nu} \frac{1}{\sqrt{(p^2 + b^2)^{\nu+\mu+1}}} \times
 \label{hyper} \\
 {}_2F_1\Bigl( \frac{\nu+\mu+1}{2}, \frac{\nu-\mu}{2}; \nu + 1;
 \frac{b^2}{p^2 + b^2} \Bigr) \nonumber
\end{eqnarray}
where the notation ${}_2F_1( a, b; c; x)$ stands for the hypergeometric
function. In many actual cases these hypergeometric functions are related
to the elementary/rational functions, since, e.g., ${}_2F_1( a, b; b; x) =
(1 - x)^{-a}$. Note also that all calculations for this study have been 
performed with the use of one- and two-electron density matrices 
\cite{MQW}, \cite{FrWa07} constructed for the incident atomic systems. 
The one-electron density matrix for three-electron Li atom is defined as 
follows 
\begin{equation}
 \rho({\bf x}_3,{\bf y}_3) = \int \int \mid \Psi( {\bf x}_1, {\bf x}_2,
 {\bf x}_3) \rangle \langle \Psi( {\bf x}_1, {\bf x}_2, {\bf y}_3) \mid
 d^3{\bf x}_1 d^3{\bf x}_2
\end{equation}
The probability to find the final one-electron atom/ion in the bound state 
with the wave function $\phi_a({\bf x}_3)$ is now written as the following 
double integral
\begin{equation}
 P = \int \int \exp(\imath {\bf V} \cdot {\bf r}_3) \phi_a({\bf x}_3) 
 \rho({\bf x}_3,{\bf y}_3) \exp(-\imath {\bf V} \cdot {\bf r}^{\prime}_3) 
 \phi_a({\bf y}_3) d^3{\bf x}_3 d^3{\bf y}_3 \label{Prob}
\end{equation}
where ${\bf x}_i = ({\bf r}_i, s_i)$ and ${\bf y}_i = ({\bf r}^{\prime}_i, 
s^{\prime}_i)$ are the spin-spatial coordinates of the $i$-th electron. It 
is assumed in Eq.(\ref{Prob}) that the bound state wave function $\phi_a$ 
is real. If the one-electron density matrix $\rho({\bf x}_3,{\bf y}_3)$ 
can be represented in a factorized form, e.g., in the form 
\begin{equation}
 \rho({\bf x}_3,{\bf y}_3) = \sum^{N}_{i=1} \sum^{N}_{j=1} C_i C_j 
 r^{n_i} \exp(-\gamma_i r_3) (r^{\prime})^{n_j} \exp(-\gamma_j 
 r^{\prime}_3) 
\end{equation}
then the formula for the final state probability is reduced to the form  
\begin{equation}
  P = \mid M_{i \rightarrow f} \mid^2 = M^{*}_{i \rightarrow f} 
 M_{i \rightarrow f}
\end{equation}
where $M_{i \rightarrow f}$ is the probability amplitude. To compute this
probability amplitude one needs to use only one-dimensional integrals. It
follows from the definition of the one-electron density matrix that 
\begin{equation}
 \int \int \rho({\bf x}_3,{\bf y}_3) d^3{\bf x}_3 d^3{\bf y}_3 = 1
\end{equation}
In actual calculations this condition is an important test of the 
constructed density matrix. Other tests were based on computation of 
different bound state properties of the Li atom with the use of 
one-electron density matrix. The two-electron density matrix is defined 
absolutely analogously and we do not want to discuss it here. The 
two-electron density matrix for the Li atom is needed in those cases 
when the goal is to evaluate the probability of formation of the He atom 
in different bound states.

The results of our computations of the final state probabilities for 
one-electron atomic species formed in the reaction Eq.(\ref{e1}) can be 
found in Table II. As mentioned above these final states include various 
ground and excited states in the helium atom (${}^{4}$He), tritium atom 
(${}^{3}$H) and He-like one-electron ion (He$^{+}$). In this study we 
restrict ourselves to the consideration of the $1s-, 2s-$ and $2p-$states 
in the final one-electron ${}^{3}$H atom and ${}^{4}$He$^{+}$ ion (see 
Table II). For the neutral ${}^4$He atom we have evaluated the final 
state probability only for its ground $1^1S-$state (see 
Eq.(\ref{heatom})). We have found that the corresponding final state 
probability is $\approx$ 7.45719$\cdot 10^{-4}$ \%, i.e. it is a 
relatively small value. Note that each nuclear fragment from reaction, 
Eq.(\ref{e1}), has a very large velocity. Therefore, it is very hard to 
observe the final atomic species with bound electrons. Based on these 
arguments one can predict that the main atomic products from reaction, 
Eq.(\ref{e1}), will be the tritium ion ${}^{3}$H$^{+}$ and helium ion 
${}^4$He$^{2+}$. The overall probability to form other atomic species is 
less than 1 \%. 

\section{Conclusion}

We have considered the nuclear reaction, Eq.(\ref{e1}), involving the ground
${}^2S-$state of the three-electron Li atom. The probabilities of formation
of different atomic species during this reaction have been evaluated
numerically, by using the actual three- and two-electron wave functions for
the Li and He atoms, respectively. Our newly developed procedure is based on 
the use of the optimized free-electron wave functions for few-electron 
atomic systems involved in the process. This allows us to perform all 
required complete and/or partial Fourier transformations of the wave 
functions. The computed final state probabilities are close to the 
exact values determined with the use of highly accurate (or completely 
correlated) wave functions which include all electron-electron (or 
correlated) coordinates. The method used in our procedure is based on the
use of one- and two-electron density matrices constructed for the ground 
state of the three-electron Li atom (for more detail, see e.g., 
\cite{FrWa07}). 

Our procedure can now be used for the more complicated nuclear reaction
Eq.(\ref{e2}) in the five-electron B-atom. The Li-atom/ion and He-atom/ion
which form during this reaction may contain up to three and two electrons,
respectively. These atomic fragments move rapidly, with the velocities
$v_{Li} \approx 2.40896$ $cm \cdot sec^{-1}$ and $v_{\alpha} \approx
4.21568$ $cm \cdot sec^{-1}$ in the case of slow neutrons. The sudden
approximation can thus certainly be applied to the He-atom and He-like ions.
However, this approximation cannot be used for internal electrons (or
$1^2s-$electrons) of the Li atom/ion, since the velocities of these two
electrons are comparable with the final velocity of the ${}^{7}$Li nucleus.
It is very likely that the probability to observe one of the 
${}^{7}$Li$^{2+}$ and ${}^{7}$Li$^{+}$ ions after reaction Eq.(\ref{e2}) 
will be relatively large  ($\ge$ 20 \%). Note also that, if the nuclear 
reaction Eq.(\ref{e1}) is produced by a fast neutron with $E_n \ge 1$ 
$MeV$, then the velocity of the Li atom/ion(s) is larger than the atomic 
velocities of the $1^2s-$electrons in the Li atom and the sudden 
approximation can be applied.

\begin{center}
    {\bf Acknowledgements}
\end{center}

It is a pleasure to thank Professor M. Frederick Hawthorne (University of
Missouri, Columbia, Missouri) for very useful references and discussion and
also the University of Western Ontario for financial support.

\newpage
  \begin{table}[tbp]
   \caption{An example of the trial, three-electron wave function
            constructed with the use of $N = 23$ semi-exponential radial
            basis functions, Eq.(7). This wave function produces the total
            energy $E$ = -7.44859276608 $a.u.$ for the ground ${}^2S-$state
            of the ${}^{\infty}$Li atom. Only one electron spin-function
            $\chi_1 = \alpha \beta \alpha - \beta \alpha \alpha$ was used
            in these calculations.}
     \begin{center}
     \scalebox{0.72}{%
     \begin{tabular}{cccccccc}
      \hline\hline
 $k$ & $m_1(k)$ & $m_2(k)$ & $m_3(k)$ & $C_{k}$ & $\alpha_k$ & $\beta_k$ & $\gamma_k$ \\
     \hline
 1  & 0 & 0 & 1 &   0.146131429481911E+02 & 0.416423958308045E+01 & 0.390330393520923E+01 & 0.754647967615054E+00 \\
 2  & 1 & 0 & 1 &   0.110681883475133E+03 & 0.526120133598745E+01 & 0.699723209902291E+01 & 0.604874234309236E+00 \\
 3  & 1 & 1 & 1 &   0.234004890152000E+03 & 0.598950989816454E+01 & 0.378246213278597E+01 & 0.590931685390717E+00 \\
 4  & 2 & 0 & 1 &   0.116882861303726E+03 & 0.413820964104499E+01 & 0.322650638788059E+01 & 0.590283203974117E+00 \\
 5  & 0 & 0 & 0 &  -0.112533836667930E+02 & 0.404234943769060E+01 & 0.402575246908481E+01 & 0.136870242306376E+01 \\
 6  & 0 & 0 & 2 &   0.124148253762594E+01 & 0.772078877480103E+01 & 0.208089611437233E+01 & 0.906692873308709E+00 \\
 7  & 0 & 0 & 3 &   0.129950444032051E+01 & 0.125634586976915E+02 & 0.167180833975207E+02 & 0.111484879880926E+01 \\
 8  & 3 & 0 & 1 &   0.121940082563610E+02 & 0.336154087416427E+01 & 0.329089537444583E+01 & 0.569461303170668E+00 \\
 9  & 2 & 2 & 1 &  -0.815264312642982E+01 & 0.315618697991587E+01 & 0.293590042487235E+01 & 0.642752252837086E+00 \\
 10 & 0 & 0 & 4 &   0.231307852206996E-01 & 0.216119418842614E+01 & 0.354528921898964E+01 & 0.848100637437686E+00 \\
 11 & 1 & 0 & 0 &  -0.290323204459308E+02 & 0.303111504960694E+01 & 0.340407714984374E+01 & 0.439831339492880E+00 \\
 12 & 1 & 0 & 2 &  -0.782805087195444E+01 & 0.319874342329211E+01 & 0.294880236447181E+01 & 0.670201325893484E+00 \\
 13 & 4 & 0 & 1 &   0.466451849833029E+05 & 0.257191421047538E+02 & 0.198922117452300E+02 & 0.745721864825035E+00 \\
 14 & 5 & 0 & 1 &   0.102312737866967E+03 & 0.610735801244370E+01 & 0.126738573321754E+02 & 0.583192334409026E+00 \\
 15 & 1 & 1 & 0 &   0.496831723313492E+01 & 0.214720891197913E+01 & 0.139151014500059E+02 & 0.332004403600092E+00 \\
 16 & 1 & 1 & 2 &   0.331570408306929E+01 & 0.313893628304852E+01 & 0.288670468153774E+01 & 0.656255394173358E+00 \\
 17 & 2 & 0 & 0 &  -0.224488896165048E+03 & 0.111494404746535E+02 & 0.712253949382151E+01 & 0.412242317869619E+00 \\
 18 & 2 & 0 & 2 &  -0.200426722868461E+01 & 0.282336074847368E+01 & 0.292360100848728E+01 & 0.638644442755579E+00 \\
 19 & 0 & 0 & 5 &   0.186252350189224E-02 & 0.103900277658347E+02 & 0.241603283627644E+01 & 0.858660133717424E+00 \\
 20 & 2 & 1 & 1 &   0.314011900712871E+01 & 0.314575640023816E+01 & 0.238785024725699E+01 & 0.645792868876097E+00 \\
 21 & 3 & 0 & 2 &   0.164525876141841E+01 & 0.318702139403640E+01 & 0.348200049008609E+01 & 0.663487519449685E+00 \\
 22 & 3 & 0 & 0 &   0.549306690646211E+01 & 0.313423005590041E+01 & 0.376361954006700E+01 & 0.494738425712773E+00 \\
 23 & 3 & 1 & 1 &   0.489167893642778E+01 & 0.310583281642480E+01 & 0.299668864388672E+01 & 0.617551331047668E+00 \\
  \hline\hline
  \end{tabular}}
  \end{center}
  \end{table}
  \begin{table}[tbp]
   \caption{The probabilities (in \%) of observing selected electron 
            final states in the helium ion ${}^4$He$^{+}$ and tritium 
            atom ${}^3$H arising in the exothermic nuclear reaction 
            Eq.(1) of the three-electron lithium-6 atom in its ground 
            ${}^2S-$state with slow neutrons.}
     \begin{center}
     \begin{tabular}{llll}
        \hline\hline
 atom/state & $1s$   &  $2s$    &  $2p$  \\
           \hline
 ${}^{4}$He$^{+}$ & 9.52386$\cdot 10^{-2}$ & 1.66605$\cdot 10^{-2}$  &  6.60170$\cdot 10^{-4}$ \\

 ${}^{3}$H & 2.49810$\cdot 10^{-3}$ & 3.05880$\cdot 10^{-4}$  &  2.00251$\cdot 10^{-6}$ \\
        \hline\hline
  \end{tabular}
  \end{center}
  \end{table}
\end{document}